\documentclass[12pt]{article}
\usepackage[utf8]{inputenc}
\usepackage{pdflscape}
\usepackage[margin=1.0in]{geometry}
\usepackage{array}
\usepackage{setspace}
\usepackage{hyperref}
\hypersetup{colorlinks=true,linkcolor=black,citecolor=blue}
\usepackage{tikz}
\usepackage{amsmath}
\usepackage{bbold}
\usepackage{slashed}
\usepackage{enumitem}
\usepackage{physics}
\usepackage{amssymb}
\usepackage{mathtools}
\usepackage{youngtab}
\usepackage{xcolor}
\usepackage{multirow}
\usepackage{stfloats}

\DeclareMathOperator{\adj}{\small\textbf{Adj}}

\DeclareMathOperator{\CS}{CS}
\DeclareMathOperator{\matter}{matter}
\DeclareMathOperator{\diag}{diag}
\usepackage{amsfonts}
\usepackage{graphicx}
\usepackage{subfiles}
\usepackage{mathrsfs}

\makeatletter
\newcommand{\thickhline}{%
    \noalign {\ifnum 0=`}\fi \hrule height 1pt
    \futurelet \reserved@a \@xhline
}
\newcolumntype{"}{@{\hskip\tabcolsep\vrule width 1pt\hskip\tabcolsep}}
\makeatother
\usepackage{imakeidx}
\makeindex[columns=3, title=Alphabetical Index, intoc]
\begin{document}
\pagenumbering{gobble}

\Large
 \begin{center}
  \textbf{Phases of $\mathbf{U(N_c)}$ QCD$_3$ from Type 0 Strings \\
   and Seiberg Duality} \\

   % Author names and affiliations
\vskip 1cm  
\large   
{Mohammad Akhond\footnote{\href{mailto:akhondmohammad@gmail.com}{akhondmohammad@gmail.com}}, Adi Armoni\footnote{\href{mailto:a.armoni@swansea.ac.uk}{a.armoni@swansea.ac.uk}} and Stefano Speziali\footnote{\href{mailto:stefano.speziali6@gmail.com}{stefano.speziali6@gmail.com}} } \\

\hspace{20pt}

\small  
\it Department of Physics,\\
Swansea University,\\
Singleton Park, Swansea,\\
SA2 8PP, U.K.
\end{center}

\hspace{20pt}

\normalsize

\begin{abstract}
  We propose an embedding of $U(N_c)$ QCD$_3$ with a Chern-Simons term in string theory. The UV gauge theory lives on the worldvolume of a Hanany-Witten brane configuration in type 0B string theory in the presence of Sagnotti's O$'3$ orientifold. We use the brane configuration to propose a magnetic Seiberg dual. We identify various phases of the magnetic theory with conjectured phases of QCD$_3$. In particular the symmetry breaking and bosonization phases are both associated with condensation of the dual squark field. We also discuss the abelian theory without Chern-Simons term and argue that flavour symmetry is not broken. Finally, we also predict novel type 0B string dynamics from QCD dynamics.         
\end{abstract}
\newpage
\pagenumbering{arabic}
%%%%%%%%%%%%%%%%%%%%%%%%%%%%%%%%
\tableofcontents
%%%%%%%%%%%%%%%%%%%%%%%%%%%%%%%%
\section{Introduction}

\paragraph{} String theory has long been a source of insight for investigations in strong coupling dynamics of quantum field theory. In particular, dualities in field theories often follow from properties of the corresponding brane configuration in string theory. Having independent evidence from field theory and string theory is a step in verifying dualities. Most of the effort so far has been largely focused on supersymmetric theories in various dimensions, owing to the fact that non-perturbative phenomena in both string theory and field theory are better understood in that setting.

One may naturally ponder the ubiquity of dualities in generic QFTs, and their relationship to string theory. Indeed, recent years have seen progress made on the field theory front for non-supersymmetric gauge theories in three dimensions. There has been significant progress in the understanding of the phase diagram of QCD$_3$ with a Chern-Simons term.

Consider a $U(N_c)$ theory with $N_f$ massless Dirac fermions and a level $K$ Chern-Simons term. It was argued \cite{Aharony:2011jz,Giombi:2011kc,Aharony:2015mjs,Hsin:2016blu} (see also \cite{Karch:2016sxi,Murugan:2016zal}) that for $N_f/2 \le K$ the theory admits a dual description in terms of a gauge theory coupled to scalars as follows
\begin{equation}\label{bosonization}
 U(N_c)_{K,K\pm N_c}\oplus N_f\;  \textnormal{fermions} \longleftrightarrow U \left(K+\frac{N_f}{2} \right)_{-N_c,-N_c\mp (K+N_f/2)}\oplus N_f\; \textnormal{scalars}\;.
\end{equation}

However, one may wonder whether something changes for $N_f/2> K$. In the case of $SU(N_c)$ gauge symmetry, it was conjectured in \cite{Komargodski-Seiberg} that when $N^\star > N_f/2 >K$ the theory admits a flavour symmetry breaking phase where
\begin{equation}\label{symmetry_breaking}
 U(N_f) \rightarrow U(N_f/2 -K) \times U(N_f/2+K) \, . 
\end{equation}
A similar picture was developed in \cite{Komargodski-Seiberg} also for $SO(N)$ and $Sp(N)$ gauge theories. For $N_f \ge N^\star$ the theory is expected to flow to a CFT\footnote{In the 't Hooft limit, when $N_c\rightarrow \infty$ and $K,N_f$ are kept fixed, the theory exhibits rich vacua \cite{Armoni:2019lgb}. The discussion of this limit is beyond the scope of this paper.}.

Following \cite{Armoni:2017jkl} which concerned the symplectic gauge group, we propose that the infrared phase diagram of $U(N_c)$ QCD$_3$ can be understood in terms of a non-SUSY Seiberg duality. Our proposal involves a modification of the UV theory, i.e. we start with a UV theory, which we refer to as the \emph{electric} theory, whose Lagrangian is more complicated than QCD$_3$. This theory flows in the IR to QCD$_3$. The electric theory also admits a Seiberg dual description, which we refer to as the \emph{magnetic} theory. The various IR phases of the electric theory (and so of QCD$_3$) can then be identified with the phases of the magnetic dual. In particular both the bosonized phase and the symmetry breaking phase, which will be our main focus, can be understood in terms of the condensation of a scalar field, namely the dual "squark", in the magnetic theory.  

Our proposal of Seiberg duality is motivated by string theory\footnote{Other approaches to obtain 3d duality with relation to string theory are given in \cite{Jensen:2017xbs,Argurio:2018uup}, while the possibility of relating these dualities to supersymmetric dualities were explored in \cite{Kachru:2016aon,Gur-Ari:2015pca}}. In order to realise $U(N_c)$ QCD$_3$ we embed the gauge theory in a Hanany-Witten brane configuration of type 0B string theory. The brane configuration consists of $N_c$ D3 branes suspended between an NS5 branes and a $(1,k)$ fivebrane. In addition, there exits $N_f$ flavour branes and an O$'3$ orientifold plane. It is similar to the corresponding supersymmetric brane configuration of Giveon and Kutasov in type IIB \cite{Giveon:2008zn}.

By swapping the fivebranes we obtain the brane configuration that realises the magnetic Seiberg dual. The relation between field theory and string theory phenomena teaches us about non-supersymmetric brane dynamics. The aforementioned squark condensation translate into a reconnection of colour and flavour branes.

Our Seiberg duality proposal passes several non-trivial checks: as in the symplectic case \cite{Armoni:2017jkl} it satisfies global anomaly matching and RG flows after mass deformations. It is also supported by planar equivalence \cite{Armoni:2003gp,Armoni:2004uu}: when $N_c,N_f,k$ are taken to infinity the electric theory becomes equivalent to a supersymmetric theory and the magnetic theory becomes equivalent to a supersymmetric theory. The electric and magnetic theories form an ${\cal N}=2$ supersymmetric Giveon-Kutasov dual pair. Therefore, there exists a limit in which our non-supersymmetric dual pair becomes a known supersymmetric dual pair.

Another method of obtaining Seiberg duality in string theory is by using non-critical strings \cite{Murthy:2006xt}. The method relies on the embedding of SQCD in non-critical string theory, pioneered in \cite{Fotopoulos:2005cn}. Instead of swapping the fivebranes, the duality is obtained by replacing the sign of the coefficient in front of the Liouville term in the string worldsheet action, $\mu \rightarrow -\mu$. The advantage of using this method is that the non-critical type 0 string does not contain a closed string tachyon in the bulk \cite{Israel:2007nj,Armoni:2008gg}. The field theory that lives on the branes is the same in both the critical and the non-critical approaches. 

In the following we will always denote the bare CS level by $k$, with $k \ge 0$. In addition, we define the frequently occurring combination
\begin{equation}\label{dictionary}
  \kappa\equiv k-N_c+2\;,\qquad  K\equiv\kappa-\frac{N_f}{2}
\end{equation}

The paper is organised as follows: in section 2 we review the essential properties of type 0B string theory and its brane configurations. In section 3 we consider a certain brane configuration and propose a Seiberg duality. In section 4 we show how the phase diagram of the electric theory manifest itself in the magnetic and in section 5 we focus on QED$_3$. Section 6 is devoted to conclusions. 

\section{Overview of type 0B}
\paragraph{}In this section we review aspects of D3 branes and O$'3$ planes in type 0 string theory. For the relevant background we refer the reader to \cite{Angelantonj:2002ct}.

Type 0B string theory can be obtained by a $\mathbb{Z}_2$ orbifold of type IIB, with the $\mathbb{Z}_2$ action generated by $(-1)^{F_s}$, the mod 2 spacetime fermion number operator. The untwisted sector is therefore identical to the bosonic sector of the parent type IIB theory. The twisted sector is composed of a tachyon in the NS-NS sector as well as a new full set of R-R fields. The tachyon will eventually be projected out by the orientifold action. The doubled set of R-R fields lead in effect to a doubling of the D-brane spectrum. In particular there are now two types of threebranes which we denote by D3 and D$3'$ respectively.

The worldvolume theory on a stack of $n$ D3 and $m$ D$3'$ branes was worked out in \cite{Blumenhagen:1999uy,Blumenhagen:1999ns}. It is a $U(n)\times U(m)$ gauge theory with 3 complex scalars in the adjoint representation, and a pair of bifundamental Weyl fermions.

In order to project out the closed string tachyon we make use of the $\Omega(-1)^{f_R}$ projection \cite{Sagnotti:1996qj,Sagnotti:1995ga}. Here, $\Omega$ is worldsheet parity and $(-1)^{f_R}$ is the operator that counts the number of right moving worldsheet fermions mod 2. Combining this with reflection in 6 spatial directions $I_6$ we get an O$'3^\pm$ orientifold, the (3+1) dimensional fixed hyperplane with respect to the $\Omega (-1)^{f_R}I_6$ action. The existence of two types of orientifold planes follows from the fact that the NS-NS two form can have a non-trivial Wilson line $\exp\left(i\int B\right)$ and the signs are chosen to reflect the R-R charge of the orientifold plane. Note that unlike the O3-planes of type IIB we do not have the additional possibilities associated with the R-R discrete torsion. Under the action of $\Omega$, D3 turns into D$3'$, thus requiring an equal number of each type of brane. In fact $\Omega$ projects out half of the doubled set of R-R fields in the closed string sector.

\begin{figure}[ht]

    \centering
    
    \begin{tikzpicture}
   \node at (0,0) {$SO(2N)$ $\mathcal{N}=4$ SYM};
      \node at (-.8,-1) {$J(-1)^F$};
   \draw[thick,->] (0,-.2)--(0,-2);
   \node at (0,-2.5) {$\mathcal{Y}^-\left({\Ylinethick{1pt}
\Yboxdim7pt\Yvcentermath1\yng(1,1)}\right)$};

 \node at (6,0) {$Sp(N)$ $\mathcal{N}=4$ SYM};
   \draw[thick,->] (6,-.2)--(6,-2);
   \node at (5.2,-1) {$J(-1)^F$};
   \node at (6,-2.5) {$\mathcal{Y}^+\left({\Ylinethick{1pt}
\Yboxdim7pt\Yvcentermath1\yng(2)}\right)$};
    \end{tikzpicture}
    \caption{The ``orientifold" daughters of $\mathcal{N}=4$ SYM.}
    \label{Orientifold projection}
\end{figure}
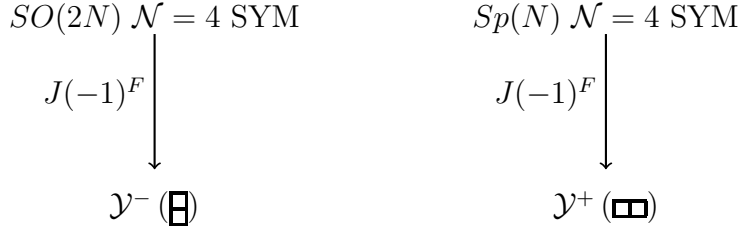

We are interested in stacks of $N$ D3 branes (together with their image $N$ D$3'$s) on top of O$'3^\pm$, with the worldvolume directions of D3 and D$3'$ parallel to that of the O$'3^\pm$-plane (see table \ref{brane orientation}). The worldvolume theory of such a configuration was worked out in \cite{Blumenhagen:1999ns}. In both cases one has a $U(N)$ gauge field and 6 adjoint scalars parameterising the directions transverse to the worldvolume. There are also a pair of Weyl fermions which transform in the 2-index symmetric or antisymmetric representation of $U(N)$ in the configuration with O$'3^+$ and O$'3^-$ respectively. We will denote these theories by $\mathcal{Y}^+\left({\Ylinethick{1pt}
\Yboxdim7pt\yng(2)}\right)$, $\mathcal{Y}^-\left({\Ylinethick{1pt}
\Yboxdim7pt\Yvcentermath1\yng(1,1)}\right)$ respectively, highlighting the orientifold type on which they live as well as the representation of the worldvolume fermions (the two features relevant for our purposes). We summarise this in table \ref{field content 4d}. The Lagrangian for these theories can be obtained by subjecting the component fields of $\mathcal{N}=4$ SYM, collectively denoted by $\varphi$, to the constraints
\begin{equation}
    J\varphi J^T=(-1)^F\varphi\;,
\end{equation}
where $(-1)^F$ is the mod 2 fermion number operator and  $J$ is the symplectic form
\begin{equation}
    J=\begin{pmatrix}0&\mathbb{1}_N\\-\mathbb{1}_N&0\end{pmatrix}\;.
\end{equation}
The choice of gauge group for the $\mathcal{N}=4$ theory descends to the choice of fermion representation (figure \ref{Orientifold projection}); starting from the \emph{parent} theory with $SO(2N)$ gauge group one lands on  $\mathcal{Y}^-\left({\Ylinethick{1pt}
\Yboxdim7pt\Yvcentermath1\yng(1,1)}\right)$, and the supersymmetric $Sp(N)$ theory leads to $\mathcal{Y}^+\left({\Ylinethick{1pt}
  \Yboxdim7pt\yng(2)}\right)$ \cite{Unsal:2006pj}.

\begin{table}[htp]

\centering
     \caption{The field content of the world volume theory of $N$ D3 branes on top of an O$'3^{\pm}$ plane.}
     \vspace{.2cm}
     \begin{minipage}{2.5in}
     \begin{tabular}{l}
    \begin{tabular}{c"c"c}

 \multicolumn{3}{c}{}\\
   
     $\mathcal{Y}^-\left({\Ylinethick{1pt}
\Yboxdim7pt\Yvcentermath1\yng(1,1)}\right)$&$U(N)$ & $SO(6)$\\\thickhline
    {$B^-_\mu$} & $\adj$ & $\cdot$  \\  
    {$X_-$} & $\adj$ & $\textbf{6}_v$  \\ 
   {$\xi_-$} & $\Ylinethick{1pt}
\Yboxdim7pt\Yvcentermath1\yng(1,1) \oplus\fontdimen8\textfont3=.75pt\overline{\tiny\Yvcentermath1\yng(1,1)}$& $\textbf{4}_s\oplus\textbf{4}_c$ \\ 
     
     \end{tabular}\\
     
    \end{tabular}
    
\end{minipage}
\begin{minipage}{2.5in}
    \begin{tabular}{c"c"c}
  \multicolumn{3}{c}{}\\
    $\mathcal{Y}^+\left({\Ylinethick{1pt}
\Yboxdim7pt\yng(2)}\right)$ &$U(N)$ & $SO(6)$ \\\thickhline
   {$B^+_\mu$} & $\adj$ & $\cdot$ \\
    {$X_+$} & $\adj$ & $\textbf{6}_v$  \\
   {$\xi_+$} & $\Ylinethick{1pt}\Yboxdim7pt\yng(2)\oplus\fontdimen8\textfont3=.75pt\overline{\tiny\yng(2)}$ & $\textbf{4}_s\oplus\textbf{4}_c$\\
     \end{tabular}
    \end{minipage}
    \label{field content 4d}
\end{table}

The M\"{o}bius amplitude for a single D3 and its image D$3'$ separated  by a distance $2|X_\pm|$ across the O$'3^\pm$ is \cite{Blumenhagen:1999ns}
\begin{equation}\label{mobius integral}
    \mathcal{A}_{\mathcal{M}}=\pm\frac{V_4}{(8\pi^2\alpha')^2}\int_0^\infty \frac{dt}{2t^3}\frac{f_2^8(iq)}{f_1^8(iq)}\exp\left({\frac{-2tX_\pm^2}{\pi \alpha'}}\right),
\end{equation}
where $q=e^{-\pi t}$ and the $f_i(q)$ are defined as in \cite{Polchinski:1996fm}. We would like to extract the charge of the orientifold plane as well as the brane-orientifold potential. We note that the integrand in \eqref{mobius integral} is, up to a sign, identical to the case analysed in \cite{Uranga:1999ib}. We will state the relevant results in the following. For large separation $X_\pm$, the leading order term as $t\xrightarrow{}0$ is given by
\begin{equation}\label{Mobius}
    \mathcal{A}_\mathcal{M}\sim\pm \pi V_4G_6(X_\pm^2)\;,
\end{equation}
where $G_6(X_\pm^2)=(4\pi^3)^{-1}|X_\pm|^{-4}\Gamma(2)$ is the 6d scalar propagator.
We see that the long range potential between the branes and O$'3^-$ (O$'3^+$) is attractive (repulsive). 
 For small $X_\pm$, \eqref{Mobius} is no longer a valid approximation, instead one can expand the exponential in \eqref{mobius integral} around $X_\pm=0$
\begin{equation}\label{mobius short}
    \mathcal{A}_\mathcal{M}=\pm\left[\Lambda-MX_\pm^2+\mathcal{O}\left(X_\pm^4\right)\right]\;,
\end{equation}
where the coefficients $\Lambda$, $M$ are both positive, with the explicit form given in \cite{Uranga:1999ib}.
From this, it follows that there is a short range attractive (repulsive) force between the branes and O$'3^-$ (O$'3^+$) plane. The nature of the interaction at short and long distances from the orientifold is similar. Therefore, the theory with fermions in the antisymmetric (symmetric) representation is perturbatively stable (unstable). Note that instabilities of non-perturbative nature may still arise, but are less straightforward to detect in string theory. Instead, we may rely on the field theory analysis and try to revert some lessons back to the brane setup (as in section \ref{region II}). 

Notice that the (in)stability of the brane configuration translates in the worldvolume field theory to statements about the vev of the scalars $X_\pm$. This is obvious from the second term in \eqref{mobius short}, where the sign of the mass term for the scalars is positive (negative) for the theory with anti-symmetric (symmetric) fermions. In the Field theory, this is encoded in the 1-loop Coleman-Weinberg potential, which gets unequal contributions from the bosons and fermions in each theory. 

As observed in \cite{Klebanov:1998yya}, the threebranes in type 0 carry the following charge and tension
\begin{equation}
Q_{D3}=\sqrt{\pi},\qquad T_{D3}=\frac{\sqrt{\pi}}{\sqrt{2}\kappa_0}\;.
\end{equation}
 It is then a matter of comparing \eqref{Mobius} with $4V_4G_6(X_\pm^2)T_{\textnormal{O$'3$}^\pm}T_{\textnormal{D3}}\kappa_0^2$ to see that the orientifold charge and tension are  
\begin{equation} \label{O3 charge}
    Q_{\textnormal{O$'3$}^\pm}=\pm\frac{Q_{\textnormal{D3}}}{2},\qquad T_{\textnormal{O$'3$}^\pm}=\pm\frac{T_{\textnormal{D3}}}{2}\;.
\end{equation}
This is clearly different from the situation in type II theories where an O$p^\pm$ plane carries $\pm2^{p-5}$ units of D$p$ brane charge. The charges \eqref{O3 charge} of the O$'3^\pm$ relative to the D3 will be crucial in constructing seiberg dual pairs in the next section.  
\subsection{A pseudo-moduli space}
\paragraph{} The discussion in the previous section shows that the $\mathcal{Y}^+\left({\Ylinethick{1pt}
\Yboxdim7pt\yng(2)}\right)$ theory is unstable, namely the D3s are repelled away from the orientifold. But the analysis tells us nothing about where the stable vacuum of the theory lies. In a non-SUSY setup, the scalar vevs, or correspondingly the coordinates of the branes are not to be viewed as moduli but are rather dictated by the dynamics of the theory. Generically one expects a scalar potential $V(X_+)$ to be induced via loop corrections. It is however useful to have a completely kinematical discussion of the possible \emph{pseudo-moduli} of the brane system before imposing the dynamical constraints. We will examine the situation both in string theory and field theory. 

Using the $U(N)$ matrices, the most generic vev for the scalars $X_+$ takes the diagonal form 
\begin{equation}\expval{X_+}=
    \diag\left(a_1,a_2,\dotsb,a_N\right);\,\qquad a_i\in \mathbb{R}\;.
\end{equation}
From a field theoretic point of view, depending on the specific values of the eigenvalues $a_i$ we encounter 3 possibilities:
\begin{enumerate}[label=(\roman*)]
    \item The $a_i$ are all distinct. In this case the gauge group is broken to its $U(1)^N$ maximal torus and the worldvolume fermions all become massive. There are also adjoint (charge 0) scalars for each $U(1)$ factor in $U(1)^N$
    \item When $n$ of the $N$ eigenvalues become exactly degenerate there is an enhanced $U(n)$ symmetry. The breaking pattern in this case takes the form
    \begin{equation}
        U(N)\rightarrow U(n)\times U(1)^{N-n}\;.
    \end{equation}
    All worldvolume fermions are massive but there are scalars in the adjoint of the unbroken gauge group. A special case of this type is when all the eigenvalues coincide and the entire gauge symmetry is unbroken.
    \item There is a more exotic possibility. Consider the situation where $n$ eigenvalues take the opposite sign of an exactly degenerate set of $m$ eigenvalues, i.e.
    \begin{equation}
        \expval{X_+}=\diag\Big(\overbrace{v,\dotsb,v}^n,\overbrace{-v,\dotsb,-v}^m,a_1,\dotsb,a_{N-(n+m)}\Big)\;.
    \end{equation}
    The unbroken gauge symmetry is now $U(n)\times U(m)\times U(1)^{N-(n+m)}$. As in the cases (i), (ii) above there are scalars transforming in the adjoint of the unbroken gauge symmetry. Unlike those cases, there are now also massless fermions thanks to the cancellation between the positive and negative eigenvalues of equal magnitude. These fermions transform in the bi-fundamental of the non-abelian $U(n)\times U(m)$ factor of the unbroken gauge group. 
\end{enumerate}

From the string theory perspective, case (i) corresponds to a configuration where all branes are at distinct points away from the orientifold, that is, none of the D3s coincide. Case (ii) corresponds to $n$ D3 branes coinciding in the bulk (away from the orientifold). Case (iii) is more interesting. Suppose that $v>0$, then in the brane picture $v$ denotes the coordinates of $n$ D3 branes in the transverse space. On the other hand giving negative vevs to $m$ of the scalars corresponds to separating $m$ D3s from the orientifold in the negative direction. But only the quotient space, i.e. the positive direction is physical. When we send $m$ D3s to a negative point in the transverse space, their image D3's are given positive coordinates and appear in the physical space. So we see that case (iii) corresponds to $n$ D3s and $m$ D3's coinciding at coordinate $v$ in the bulk. The worldvolume theory of this configuration beautifully matches what one would expect from field theory discussed in (iii).
 
\subsection{Hanany-Witten setup}

\begin{table}[ht]
    \centering
     \caption{The various extended objects and their orientation in $\mathbb{R}^{1,9}$. All objects extend along the shared $x^{0,1,2}$ directions as well as those indicated below.}
     \vspace{.5 cm}
    \begin{tabular}{|c"ccccccc|}\thickhline
        NS5 & 3 & 4 & 5 &&&&\\\thickhline
        NS$5'$ & 3 &&&&&8&9\\\thickhline
        D3 &  &  &  &$|6|$&&&\\\thickhline
        O$'3$ &  &  &  &6&&&\\\thickhline
        D5 &  &  &  &&7&8&9\\\thickhline
        $(1,k)$&$\tiny\begin{bmatrix}3\\7\end{bmatrix}_{\theta}$&&&&&8&9\\\thickhline
    \end{tabular}
    \label{brane orientation}
\end{table}

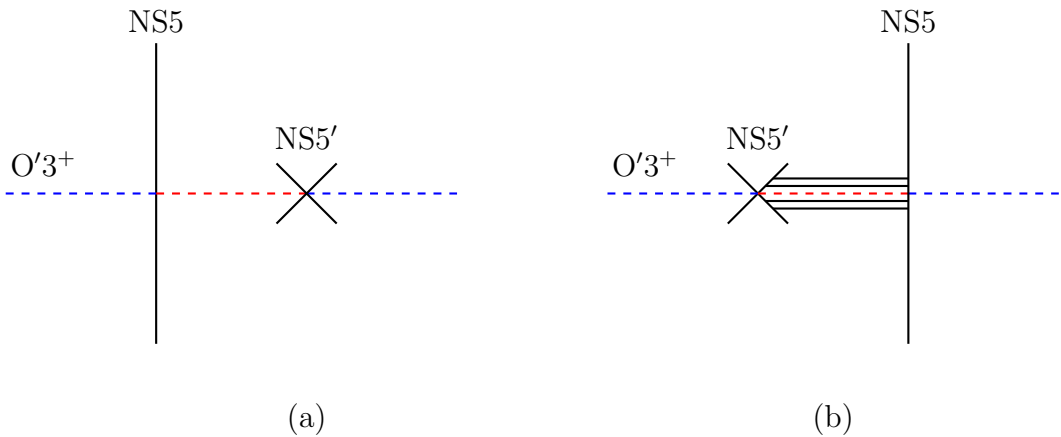
\begin{figure}[ht]
    \centering
    
    \begin{tikzpicture}
    \draw[thick] (0,-2)--(0,2) node[anchor=south]{NS5};
    \draw[thick,dashed, blue] (-2,0)--(0,0);
    \draw[thick,dashed,red] (0,0)--(2,0);
     \draw[thick,dashed, blue] (2,0)--(4,0);
    \draw [thick] (2,0)--(2.4,.4);
    \draw[thick] (2,0)--(1.6,-.4);
    \draw[thick] (2,0)--(2.4,-.4);
    \draw[thick] (2,0)--(1.6,.4);
    \node at (2,.75) {NS$5'$};
    \node at (-1.5,.4) {O$'3^+$};
  
    \draw [thick] (8,0)--(8.4,.4);
    \draw[thick] (8,0)--(7.6,-.4);
    \draw[thick] (8,0)--(8.4,-.4);
    \draw[thick] (8,0)--(7.6,.4);
    \draw[thick,dashed, blue] (6,0)--(8,0);
    \draw[thick,dashed,red] (8,0)--(10,0);
     \draw[thick,dashed, blue] (10,0)--(12,0);
    \draw[thick] (8.1,.1)--(10,0.1);
    \draw[thick] (8.2,.2)--(10,0.2);
    \draw[thick] (8.1,-0.1)--(10,-0.1);
    \draw[thick] (8.2,-.2)--(10,-0.2);
    \draw [thick] (10,-2)--(10,2) node[anchor=south]{NS5};
    \node at (8,.75) {NS$5'$};
    \node at (2,-3) {(a)};
    \node at (9,-3) {(b)};
      \node at (6.5,.4) {O$'3^+$};
    
    \end{tikzpicture}
    \caption{The Hanany-Witten effect. In passing from the configuration (a) to (b) a pair of D3s are created between the non-parallel NS5s. }
    \label{HW transition}
\end{figure}

\paragraph{} We are interested in Hanany-Witten setups to study 3d theories, which requires the introduction of NS5 branes. Our construction is the non-SUSY analogue of the 3d $\mathcal{N}=2$ setup in type IIB (see e.g. \cite{Giveon:1998sr}). In particular, we have NS5 branes which are non-parallel in two of their spatial coordinates as in table \ref{brane orientation}, we distinguish them by referring to one as an NS$5'$. The orientifold charge is switched from O$'3^+$ to O$'3^-$ and vice versa on either side of an NS5 or NS$5'$ which intersects the orientifold. We will only consider configurations where the orientifold is asymptotically O$'3^+$ and label only the asymptotic charge of the orientifold plane in our diagrams (see figure \ref{HW transition}).  

Seiberg duality has a standard string theory derivation \cite{Elitzur-Giveon-Kutasov} which follows from a rearrangement of non-parallel NS5 branes in the Hanany-Witten setup. In constructions without an orientifold, it is possible to achieve this rearrangement without the need for the NS5 branes to intersect. This is done by using the freedom to separate them in a direction mutually transverse to the NS5 and NS$5'$. In the presence of an orientifold, the NS5s are bound to the orientifold plane and this is no longer possible. The NS5 branes will inevitably intersect as we try to move them past one another \cite{Evans:1997hk}.

The result of moving non-parallel fivebranes through one another in the presence of an orientifold is well understood. This is the so called Hanany-Witten transition \cite{Hanany-witten}. In type IIB constructions with an orientifold this amounts to the creation/annihilation of a D3 between the NS5 and NS$5'$ depending on the orientifold type, a fact that follows from imposing the conservation of linking number. In the absence of D5 branes the linking number of an NS5 is proportional to the difference of the net D3 brane charges ending on it from the left and right respectively. Following the discussion around \eqref{O3 charge} it is easy to see that for the type 0 configuration of figure \ref{HW transition} the linking number of the NS5 and NS$5'$ are conserved provided a \emph{pair} of D3s are created in between them as we go from (a) to (b). This is twice the corresponding situation in type IIB as one would expect from the fact that the charge of O$'3^\pm$ relative to the type 0 D3 is a factor of two greater than the type IIB analogue.

In the next section we discuss the Hanany-Witten setup that leads to the non-SUSY gauge theories of interest with and without flavours.

 \section{3d dualities from non-supersymmetric brane configurations}

In this section we consider Hanany-Witten setups that lead to three-dimensional CS theories. See figure \ref{level-rank brane diagram} and \ref{flavoured brane diagram}. The construction is analogous to \cite{Giveon:1998sr}. The difference here, besides being in type 0B, is the presence of the O$'3$ orientifold discussed previously.

In section \ref{Level-rank duality} we consider the setup of figure \ref{level-rank brane diagram}. The low-energy theory of such a configuration is that of non-SUSY analogue of $\mathcal{N} = 2$ CS theories without flavours of (s)quarks. Such a setup turns out to be meaningful for the discussion of 3d dualities without matter. These dualities are also known in the literature as level-rank dualities.

In section \ref{Including flavours} we consider the addition of $N_f$ flavour D5-branes, see figure \ref{flavoured brane diagram}. The low-energy theory emerging from such a brane configuration includes quarks and squarks in the fundamental representation of the gauge group.
 
  \subsection{Level-rank duality}\label{Level-rank duality}
  \paragraph{} We begin by discussing how level-rank duality is realised in our setup. The discussion follows that of \cite{Ireson:2014cia}, and we provide a more refined account. In particular, we will be more careful about the CS level of the $U(1)$ factor of the gauge group. 

\begin{figure}[t]
    \centering
    
    \begin{tikzpicture}
    \draw[thick] (0,-2)--(0,2) node[anchor=south]{NS5};
    \draw[thick,dashed, blue] (-2,0)--(0,0);
    \draw[thick,dashed,red] (0,0)--(2,0);
    \node[label=below:{$(1,k)$}] at (2,-0.1){};
    \draw (0,.1)--(2,0.1);
    \draw (0,-0.1)--(2,-0.1);
    \node[label=above:{$N_c$ D3}] at (1,0.1){} ;
    \draw[thick,blue,dashed](2,0)--(4,0) ;
    \draw [black, fill] (2,0) circle [radius=.2];
    \node at (-1.5,0.4) {O$'3^+$};

    \draw[thick,dashed,blue] (7,0)--(9,0);
    \draw[thick,dashed,red] (9,0)--(11,0);
    \draw[thick,dashed,blue] (11,0)--(13,0);
    \draw (9,.1)--(11,.1);
    \draw (9,-.1)--(11,-.1);
    \draw[thick](11,-2)--(11,2) node[anchor=south]{NS5};
    \node[label=above:{$\kappa$ D3}] at (10,.1){};
    \node[label=below:{$(1,k)$}] at (9,-.1){};
    \draw [black, fill] (9,0) circle [radius=.2];
    \node at (7.5,0.4) {O$'3^+$}; 
    
    \node at (1,-3) {(a)};
    \node at (10,-3) {(b)};
    \end{tikzpicture}
    \caption{The brane setup for the (a) electric and (b) magnetic theory which give rise to level-rank duality.}
    \label{level-rank brane diagram}
\end{figure}
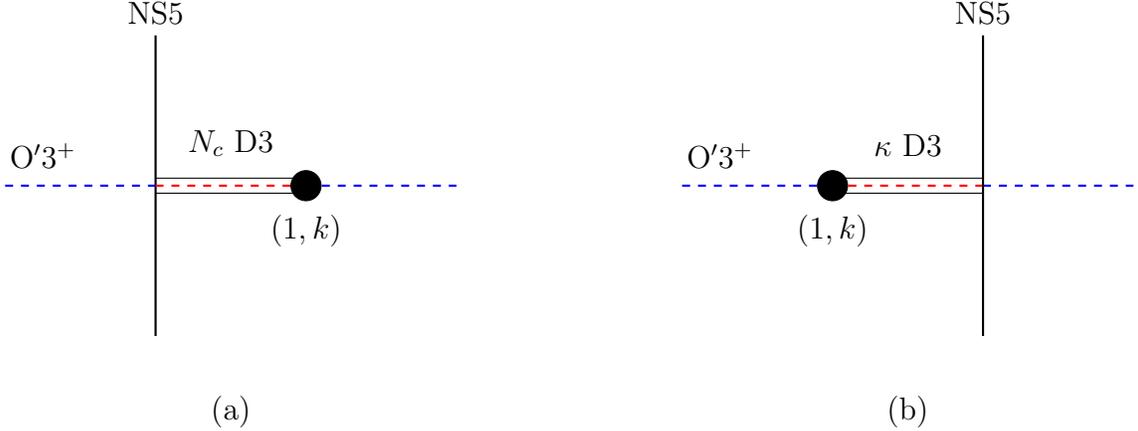

The starting point is the brane configuration (a) of figure \ref{level-rank brane diagram} with $N_c$ D3 branes stretched between an NS5 brane and a $(1,k)$ 5-brane. We will refer to this as the \emph{electric} theory. The worldvolume theory is the dimensional reduction of the $\mathcal{Y}^-\left({\Ylinethick{1pt}
\Yboxdim7pt\Yvcentermath1\yng(1,1)}\right)$ subject to suitable boundary conditions. There is a $U(N_c)$ gauge field $A_\mu$ with a YM term and level $k$ CS interactions, as well as a real scalar $\sigma$ in the adjoint of $U(N_c)$ and two antisymmetric (complex) Dirac fermions in the $\Ylinethick{1pt}\Yboxdim7pt\Yvcentermath1\yng(1,1)$ and the ${\overline{\Ylinethick{1pt}
\Yboxdim7pt\Yvcentermath1\yng(1,1)}}$ of $U(N_c)$, respectively. The Lagrangian takes the following form\footnote{Such a Lagrangian is understood as descending from its parent $\mathcal{N} = 2$ counterpart. In the large $N$ limit we expect to recover a supersymmetric YM-CS theory. The following rule is expected to hold: $\Ylinethick{1pt}
\Yboxdim7pt\Yvcentermath1\yng(1,1) \oplus\fontdimen8\textfont3=.75pt\overline{\Yboxdim7pt\Yvcentermath1\yng(1,1)} \rightarrow$ \textbf{Adj}.}
\begin{equation}\label{Electric Lagrangian gauge}
\begin{split}
\mathscr{L}^{(E)}_{N_f=0} = &\frac{1}{g_e^2}\Tr\left[-\frac{1}{2}(F_{\mu\nu})^2+(D_\mu\sigma)^2+i\Bar{\lambda}\slashed{D}\lambda + i \Bar{\tilde{\lambda}} \slashed{D} \tilde{\lambda} - i \Bar{\lambda}\sigma\lambda - i \Bar{\tilde{\lambda}} \sigma \tilde{\lambda} + D^2 \right]\\
&+\frac{k}{4\pi}\Tr\left[ \epsilon^{\mu\nu\rho}\left(A_\mu \partial_\nu A_\rho-\frac{2i}{3}A_\mu A_\nu A_\rho \right) + 2 D \sigma - \Bar{\lambda}\lambda - \Bar{\tilde{\lambda}} \tilde{\lambda}  \right] \, .
\end{split}
\end{equation}
Here $F_{\mu\nu}$ is the gauge field strength and $D_\mu\equiv \partial_\mu-i A_\mu$ is the covariant derivative. The covariant derivative is understood to act on the various fields in the representations of $U(N_c)$ they belong to. $D$ is the auxiliary field of the vector multiplet borrowed from the supersymmetric parent theory. It belongs to the adjoint representation of the gauge group just like the gauge field and scalar gaugino.

\begin{table}[htp]
\centering
     \caption{The field content of the worldvolume theories of the brane constructions in figure \ref{level-rank brane diagram}}
     \vspace{.2cm}
     \begin{minipage}{2in}
    
    \begin{tabular}{c"c}

 \multicolumn{2}{c}{}\\
     &$U(N_c)_{k}$\\\thickhline 
   {$A_\mu$} & $\adj$  \\
   {$\sigma$} & $\adj$  \\ 
   {$\lambda$} & $\Ylinethick{1 pt}\Yboxdim7pt\yng(1,1)$\\
   {$\tilde{\lambda}$} & $\overline{\Ylinethick{1 pt}\Yboxdim7pt\yng(1,1)}$\\
     \end{tabular}
     
\end{minipage}
\begin{minipage}{2in}
    \begin{tabular}{c"c}

 \multicolumn{2}{c}{}\\
     &$U(\kappa)_{-k}$\\\thickhline 
   {$a_\mu$} & $\adj$  \\ 
   {$s$} & $\adj$  \\  
   {$l$} & $\Ylinethick{1 pt}\Yboxdim7pt\yng(1,1)$\\ 
   {$\tilde{l}$} & $\overline{\Ylinethick{1 pt}\Yboxdim7pt\yng(1,1)}$\\
     
     \end{tabular}

    \end{minipage}
    \label{field content level-rank}
\end{table}
It is straightforward to obtain the Seiberg dual of this theory following e.g. \cite{Evans:1997hk,Giveon:1998sr} with a slight modification that takes into account the effect discussed in figure \ref{HW transition}. After reshuffling the NS5 and $(1,k)$ fivebrane we arrive at the configuration (b) in figure \ref{level-rank brane diagram}, where the number of colour D3s is now $\kappa\equiv k-N_c+2$. We refer to this as the \emph{magnetic} theory. The worldvolume theory is now that of a gauge field $a_\mu$ with YM term and level $-k$ CS interactions as well as a real adjoint scalar $s$ and antisymmetric Dirac fermions $l$ and $\tilde{l}$. The Lagrangian is
\begin{equation}\label{magnetic lagrangian YM}
\begin{split}
\mathscr{L}_{N_f=0}^{(M)} = &\frac{1}{g_m^2}\Tr\left[-\frac{1}{2}(f_{\mu\nu})^2+(D_\mu s)^2+i\Bar{l}\slashed{D}l + i \Bar{\tilde{l}} \slashed{D} \tilde{l} - i \Bar{l} s l - i \Bar{\tilde{l}} s \tilde{l} + D^2 \right]\\
&+\frac{k}{4\pi}\Tr\left[ \epsilon^{\mu\nu\rho}\left(a_\mu \partial_\nu a_\rho-\frac{2i}{3} a_\mu a_\nu a_\rho \right) + 2 D s - \Bar{l} l - \Bar{\tilde{l}} \tilde{l}  \right] \, .
\end{split}
\end{equation}

We are interested in the IR dynamics of these theories. In the absence of supersymmetry, the scalars on the two sides are expected to acquire a 1-loop mass of the order of the cutoff \cite{Ireson:2014cia}
\begin{equation}\label{scalar gaugino mass}
    m^2_\sigma\sim g_e^2\Lambda,\quad  m^2_s\sim g_m^2\Lambda \;.
\end{equation}
As in the discussion following \eqref{mobius short} this translates to an attractive force between the branes and the orientifolds, signalling perturbative stability of the configuration. At energies well below the cutoff scales, the scalars are decoupled and do not play a role. Note that the scalars also have tree level CS masses, but we expect them to be subleading due to the stringy nature of the masses in \eqref{scalar gaugino mass}. After integrating out the scalars we are left with gauge fields and antisymmetric fermions, both of which have tree-level CS masses $M_{CS}=\pm g^2k$ where the sign of the mass follows from the sign of the bare CS levels in \eqref{Electric Lagrangian gauge} and \eqref{magnetic lagrangian YM}. Due to the lack of supersymmetry, also the gauginos (the antisymmetric fermions) get a mass at one-loop and can be integrated out. Integrating out the antisymmetric fermions shift the levels of the $U(1)$ and $SU(N_c)$ (resp. $SU(\kappa)$) factors of the gauge group by disproportionate amounts. As a result the IR of the electric theory is a $U(N_c)_{K_1,K_2}$ CS TQFT where
\begin{equation}\label{gaugino level-shift electric}
    K_1=k-N_c+2\equiv \kappa,\qquad K_2=k-2N_c+2\equiv \kappa -N_c\;. 
\end{equation}
While the IR of the magnetic theory is described by a $U(\kappa)_{L_1,L_2}$ CS TQFT with
\begin{equation}\label{gaugino level-shift magnetic}
    L_1=-k+\kappa-2=-N_c\;,\qquad L_2=-k+2\kappa-2=-N_c+\kappa\;.
\end{equation}
Putting everything together we end up with the TQFTs $U(N_c)_{\kappa,\kappa-N_c}$ and $U(\kappa)_{-N_c,-N_c+\kappa}$, In fact, these theories are dual to each other. Therefore, in the IR, we recover the following level-rank duality
\begin{equation}
    U(N_c)_{\kappa,\kappa-N_c}\longleftrightarrow U(\kappa)_{-N_c,-N_c+\kappa}\;.
\end{equation}
 \subsection{Including flavours}\label{Including flavours}
 \paragraph{} We can include flavours in the discussion by adding D5 branes to the setup, the worldvolume directions spanned by the flavour D5 branes are as in table \ref{brane orientation}. The IR phases of the electric theory turn out to be richer than the cases studied above and are nicely encoded in terms of the dual magnetic theory. We begin by analysing each theory separately semi-classically before mapping out the phase diagram.

\begin{figure}[t]
    \centering
    
    \begin{tikzpicture}[scale=.95]
    \draw[thick] (0,-2)--(0,2) node[anchor=south]{NS5};
    \draw[thick,dashed, blue] (-2,0)--(0,0);
    \draw[thick,dashed,red] (0,0)--(4,0);
    \node[label=above:{$(1,k)$}] at (4,0){};
    \draw (0,.1)--(4,0.1);
    \draw (0,-0.1)--(4,-0.1);
    \node[label=above:{$N_c$ D3}] at (1,0.1) {};
    \node[label=below:{$N_f$ D5}] at (2,-0.1){};
    \draw[thick,blue,dashed](4,0)--(6,0) ;
    \draw [black, fill] (4,0) circle [radius=.2];
    \draw [black, fill] (1.85,-0.15) rectangle (2.15,0.15);
    \node at (-1.5,.4) {O$'3^+$};
    
    \draw[thick,dashed,blue] (7,0)--(9,0);
    \draw[thick,dashed,blue] (9,0)--(11,0);
    \draw[thick,dashed,red] (11,0)--(13,0);
    \draw[thick,dashed,blue] (13,0)--(15,0);
    \draw (9,.1)--(13,.1);
    \draw (9,-.1)--(13,-.1);
    \draw[thick](13,-2)--(13,2) node[anchor=south]{NS5};
    \node[label=above:{$N_f$ D5}] at (9,.1){};
    \node[label=above:{$\tilde{N_c}$ D3}] at (12,.1){};
    \node[label=below:{$(1,k)$}] at (11,-.1){};
    \draw [black, fill] (11,0) circle [radius=.2];
    \draw [black, fill] (8.85,-0.15) rectangle (9.15,0.15) ;
    \node at (7.5,.4) {O$'3^+$};
    \node at (2,-3) {(a)};
    \node at (11,-3) {(b)};
    \end{tikzpicture}
    \caption{The brane setup for the (a) electric and (b) magnetic theory of our proposal. Here $\tilde{N_c}=N_f+k+2-N_c$}
    \label{flavoured brane diagram}
\end{figure}
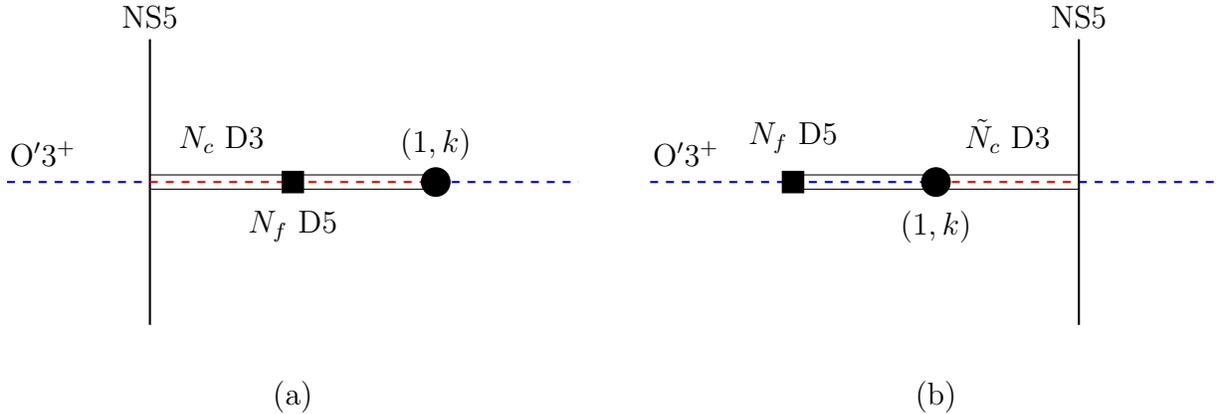

\begin{table}[ht]
\centering
     \caption{The field content of the electric and magnetic theory.}
     \vspace{.5cm}
     \begin{minipage}{2.5in}
     \begin{tabular}{l}
    \begin{tabular}{c"c"c}
     \\[-5em]
  \thickhline
 
  \multicolumn{3}{c}{Electric Theory}\\\thickhline 
 \multicolumn{3}{c}{}\\

     &$U(N_c)_{k}$ & $SU(N_f)$ \\ \thickhline 
   {$A_\mu$} & $\adj$ & $\cdot$  \\
   {$\sigma$} & $\adj$ & $\cdot$  \\ 
   {$\lambda$} & $\Ylinethick{1 pt}\Yboxdim7pt\yng(1,1)$ & $\cdot$  \\ 
   {$\tilde{\lambda}$} & $\overline{\Ylinethick{1 pt}\Yboxdim7pt\yng(1,1)}$ & $\cdot$  \\ \thickhline 
  {$\Phi$} & $\overline{\Ylinethick{1 pt}\Yboxdim7pt\Yvcentermath1\yng(1)}$ & $\Ylinethick{1 pt}\Yboxdim7pt\Yvcentermath1\yng(1)$ \\
     {$\Psi$} & $\Ylinethick{1 pt}\Yboxdim7pt\Yvcentermath1\yng(1)$ & $\Ylinethick{1 pt}\Yboxdim7pt\Yvcentermath1\yng(1)$ \\
     
     \end{tabular}\\
     
    \end{tabular}
    
\end{minipage}
\begin{minipage}{2.5in}
    \begin{tabular}{c"c"c}
  \thickhline
  \multicolumn{3}{c}{Magnetic Theory}\\\thickhline 
  \multicolumn{3}{c}{}\\
  
     &$U(\tilde{N_c})_{-k}$ & $SU(N_f)$ \\ \thickhline
   {$a_\mu$} & $\adj$ & $\cdot$ \\ 
   {$s$} & $\adj$ & $\cdot$ \\
   {$l$} & $\Ylinethick{1 pt}\Yboxdim7pt\yng(1,1)$ & $\cdot$ \\
   {$\tilde{l}$} & $\overline{\Ylinethick{1 pt}\Yboxdim7pt\yng(1,1)}$ & $\cdot$ \\\thickhline 
  {$\phi$} & $\Ylinethick{1 pt}\Yboxdim7pt\Yvcentermath1\yng(1)$ & $\overline{\Ylinethick{1 pt}\Yboxdim7pt\Yvcentermath1\yng(1)}$ \\ {$\psi$} & $\overline{\Ylinethick{1 pt}\Yboxdim7pt\Yvcentermath1\yng(1)}$ & $\overline{\Ylinethick{1 pt}\Yboxdim7pt\Yvcentermath1\yng(1)}$ \\ \thickhline 
     {$M$} & $\cdot$ & $\adj$ \\
     {$\chi$} & $\cdot$ & $\Ylinethick{1 pt}\Yboxdim7pt\yng(2)$ \\
     {$\tilde{\chi}$} & $\cdot$ & $\overline{\Ylinethick{1 pt}\Yboxdim7pt\yng(2)}$ \\ 
     \end{tabular}
    \end{minipage}
    \label{field content}
\end{table}

\subsubsection{Electric theory}
\paragraph{} The flavoured electric theory is realised on the brane configuration (a) of figure \ref{flavoured brane diagram}. The worldvolume theory on the D3 branes now includes $N_f$ complex scalars $\Phi$ and $N_f$ Dirac fermions $\Psi$. The relevant flavour symmetry emerging from the branes is an $SU(N_f)$ group. The representations of the scalars and fermions with respect to the gauge and flavour groups are listed in table \ref{field content}. These are essentially determined by their coupling to the antisymmetric gauginos, see later \eqref{electric lagrangian matter}.

The tree level Lagrangian is given by
\begin{equation}
    \mathscr{L}^{(E)}=\mathscr{L}^{(E)}_{N_f = 0} + \mathscr{L}_{\matter}\;,
\end{equation}
where $\mathscr{L}^{(E)}_{N_f = 0} $ is, as before, given by \eqref{Electric Lagrangian gauge}. The additional flavour terms are described by
\begin{equation}\label{electric lagrangian matter}
\begin{split}
\mathscr{L}_{\matter} = &\lvert D_\mu\Phi^a_i \rvert^2 + i\Bar{\Psi}^{ai} (\slashed{D}\Psi)_{a i} -\Bar{\Phi}^i_a{(\sigma^2)}^a_b \Phi^{b}_i + \Bar{\Phi}^i_a{(D^2)}^a_b \Phi^{b}_i  \\
    & - \Psi_{a i} \sigma_b^a \bar{\Psi}^{b i} - ( i{\lambda_{[ab]}} \Phi^a_{i} \bar{\Psi}^{bi}   + i{\tilde{\lambda}^{[ab]}} \bar{\Phi}^i_{a} \Psi_{bi}  + \text{h.c.})\;.
\end{split}
\end{equation}
Here $a,b=1,\dotsb,N_c$ are colour indices and $i,j=1,\dotsb,N_f$ are flavour indices. The interactions with the gauginos fix the representations of the (s)quark fields to be as in table \ref{field content}.

The fate of the scalar $\sigma$ of the gauge multiplet of the electric theory is similar to the flavourless case. The one-loop corrections to the scalar propagator get positive contributions from its coupling to itself and to the gauge field and negative contributions from its coupling to the gaugino $\lambda$. Since there are more bosonic than fermionic degrees of freedom, the vacuum $\expval{\sigma}=0$ is stable; $\sigma$ does not play a role in the IR dynamics of the theory and can be integrated out.

A similar story pans out for the squark $\Phi$. Indeed, the squark couples to the gauge field $A_\mu$, the scalar $\sigma$ and the gaugino $\lambda$. Since there are more bosonic than fermionic degrees of freedom, one expects the squark to acquire a positive mass $M_\Phi^2>0$ and decouple from the IR physics.

For a non-zero level $k\neq 0$, the gauge field and the gaugino acquire a Chern-Simons mass $M_{\CS}=g^2k$. We therefore expect the IR physics to be dominated by the topological CS theory coupled to $N_f$ fundamental quarks, i.e. QCD$_3$ with $N_f$ quark flavours.\footnote{Integrating out the gauge sector is somewhat more natural in the semiclassical regime $k\gg1$. We expect this to remain true also at finite $k$, unless something drastic happens.} The IR levels of the electric theory are shifted by the gaugino as in \eqref{gaugino level-shift electric}, as well as the fundamental quarks. In summary, using the dictionary \eqref{dictionary} we have
\begin{equation}\label{electric IR}
    \text{electric IR:}\qquad U(N_c)_{K,K-N_c}\oplus N_f\;\text{fermions}\;,
\end{equation}
which is nothing but the left hand side of \eqref{bosonization}.

On the other hand, when $k=0$, the IR theory is that of YM theory coupled to the gaugino and the fundamental quarks. It is less straightforward to say anything concrete about the IR dynamics of this theory.

\subsubsection{Magnetic theory}
\paragraph{} The flavoured magnetic theory lives on the configuration (b) of figure \ref{flavoured brane diagram}. It is obtained from the flavoured electric theory by the standard Giveon-Kutasov move \cite{Giveon:1998sr,Evans:1997hk} modified so as to account for the brane creation described in figure \ref{HW transition}. One can easily verify that the resulting number of colour branes between the NS5 and the $(1,k)$ fivebrane is 
\begin{equation}
\tilde{N}_c=N_f+k-N_c+2\equiv N_f+\kappa\;.
\end{equation}

The magnetic field content is given in table \ref{field content}. This can be obtained in a similar fashion to the electric theory, i.e. by subjecting the theory on the D3 branes in table \ref{field content 4d} to the appropriate boundary conditions. We have a gauge multiplet identical to the magnetic theory of the $N_f=0$ case. The matter multiplet consists of a complex scalar $\phi$ and a Dirac fermion $\psi$. Their representations with respect to the gauge and flavour groups are given in table \ref{field content}. There are in addition new degrees of freedom, which have no analogue on the electric side, corresponding to the motion of the flavour D3 branes along the $x^{8,9}$ directions. These give rise to two gauge singlets; the meson $M$ which is an $SU(N_f)$ adjoint and its fermionic partners, the ``mesinos" $\chi$ transforming as $\Ylinethick{1 pt}\Yboxdim7pt\Yvcentermath1\yng(2)$ of $SU(N_f)$ and $\tilde{\chi}$ transforming as $\overline{\Ylinethick{1 pt}\Yboxdim7pt\Yvcentermath1\yng(2)}$ of $SU(N_f)$.

The tree level Lagrangian for this theory is 
\begin{equation}
    \mathscr{L}^{(M)}=\mathscr{L}^{(M)}_{N_f = 0} + \mathscr{L}_{\matter}\;,
\end{equation}
where $\mathscr{L}^{(M)}_{N_f = 0}$ is as in \eqref{magnetic lagrangian YM}. The matter Lagrangian is
\begin{equation}\label{magnetic lagrangian matter}
\begin{split}
\mathscr{L}_{\matter} = &\lvert D_\mu\phi^{i}_a \rvert^2 + i \bar{\psi} (\slashed{D} \psi)^{ai} - \bar{\phi}^a_i (s^2)^{b}_a \phi^i_b + \bar{\phi}^a_i D^{b}_a \phi_{bi} - \psi^{a i} (s)^{b}_a \bar{\psi}_{bi} \\
& - \left( i \tilde{l}^{[ab]} \phi_a^i \bar{\psi}_{bi} + i l_{[ab]} \bar{\phi}_i^a \psi^{bi} + \text{h.c.} \right) + \lvert \partial_\mu M^{i}_j \rvert^2 + i \bar{\chi}^{ \{ i j \} } \slashed{\partial}\chi_{ \{ ij \}} \\
& - y^2 \bar{\phi}^a_i \phi_a^i \bar{\phi}^b_j \phi_b^j - y^2 \phi_a^i \bar{M}^j_i M^k_j \bar{\phi}_k^a - y \Big(\chi _{ \{ i j \}} \phi_a^i \psi^{aj} + \tilde{\chi}^{ \{ i j \}} \bar{\phi}^a_i \bar{\psi}_{aj}  + \text{h.c.} \Big) \\
&- y \left(\psi^{a i} M^j_i \bar{\psi}_{a j}+\text{h.c.}\right) \, .
\end{split}
\end{equation}
Note that in addition to the magnetic gauge coupling $g_m$, we now have another coupling constant $y$ which controls interactions between the (s)quarks and the meson multiplet.

The scalar $s$ of the magnetic gauge multiplet gets a positive mass and decouples, just as it did in the flavourless case. This signals the stability of the colour branes near the orientifold.

The squark $\phi$ couples to the gauge multiplet as well as the meson multiplet. There are more bosonic than fermionic degrees of freedom in the gauge multiplet, and more fermionic than bosonic degrees of freedom in the meson multiplet. Therefore, the squark aquires a 1-loop mass of the form
\begin{equation} \label{magnetic squark mass}
    M_\phi^2\sim (-y^2+g_m^2)\Lambda \; .
\end{equation}
The two effects compete and the squark may become massive or tachyonic. Since at large $k$ the gauge field becomes heavy and decouples we operate under the assumption that in this limit the squark is tachyonic.

The matter Lagrangian \eqref{magnetic lagrangian matter} for the magnetic theory includes a coupling between the meson field and the scalar quarks
\begin{equation}
    y^2 \phi_a^i \bar{M}_i^j M_j^k \bar{\phi}_k^a   \;.
\label{phiMterm}
\end{equation}
If the meson acquires a vev of the form $\expval{\bar{M}_{i}^j M_{j}^k} = u^2\delta^k_i$ the squark $\phi$ becomes massive. If the squark acquires a vev $\expval{\phi^i_a}=v \delta_a^i$, and flavour symmetry is unbroken, the mesons become massive. Therefore, the most likely scenario is that in all phases \cite{Armoni:2017jkl}
\begin{equation}
    M_\phi^2M_M^2<0\;.
\end{equation}
In the following we will always work with this assumption in mind. This will be crucial in obtaining the phase diagram of QCD$_{3}$. 
\section{Phase diagram}
\paragraph{} As we saw in \eqref{electric IR}, the IR theory on the electric brane configuration is precisely QCD$_3$. In this section we argue that the conjectured phase diagram of QCD$_3$ can be understood in terms of the dual magnetic description. Many of the features are similar to the symplectic case analysed in \cite{Armoni:2017jkl}. For this reason we will be somewhat brief and focus only on the details which are new to the unitary theory.
\subsection{Region I: Bosonization}
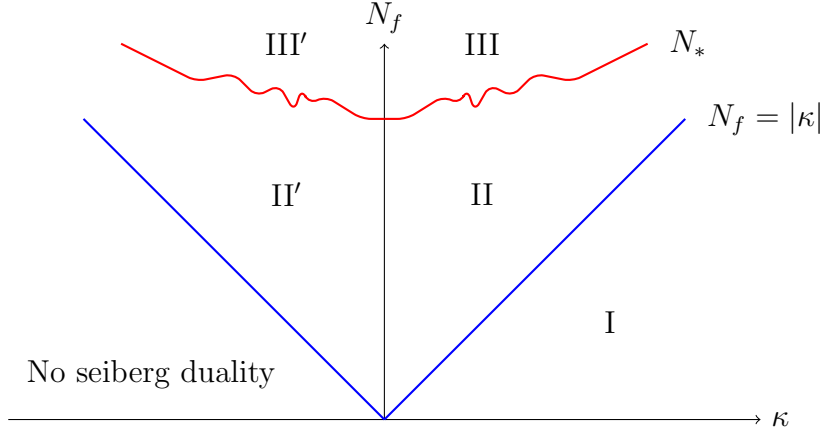
\begin{figure}[t]
    \centering
    \begin{tikzpicture}[domain=0:4]
    \draw (-5,0)--(0,0);
    \draw[->](0,0)--(5,0) node[right] {$\kappa$};
    \draw[->](0,0)--(0,5) node[above] {$N_f$} ;
    \draw[thick, color=blue]    plot (\x,\x);
    \node[label=right:{$N_f=|\kappa|$}] at (4,4) {};
    \node[label=right:{$N_*$}] at (3.5,5) {};
    \draw[thick, color=blue]    plot ({-\x},\x);
    \draw[thick,color=red,rounded corners=3.4pt] (0,4)--(.3,4)--(.78,4.3)--(1,4.2)--(1.1,4.4)--(1.2,4.1)--(1.4,4.45)--(1.7,4.3)--(2,4.6)--(2.5,4.5)--(3.5,5);
     \draw[thick,color=red,rounded corners=3.4pt] (0,4)--(-.3,4)--(-.78,4.3)--(-1,4.2)--(-1.1,4.4)--(-1.2,4.1)--(-1.4,4.45)--(-1.7,4.3)--(-2,4.6)--(-2.5,4.5)--(-3.5,5);
     \node at (3,1.3) {I};
     \node at (1.3,3) {II};
     \node at (-1.3,3) {II$'$};
     \node at (-3.1,.6) {No seiberg duality};
     \node at (1.3,5) {III};
     \node at (-1.3,5) {III$'$};
    \end{tikzpicture}
    \caption{Phase diagram of $\textnormal{QCD}_3$.} 
    \label{phase diagram}
\end{figure}

\paragraph{} We start with the region of the parameter space where $\kappa\equiv k+2-N_c\geq N_f$. This corresponds to region I in the phase diagram of figure \ref{phase diagram}. In this region the rank of the magnetic gauge group $\tilde{N}_c=N_f+\kappa$ is automatically positive. Following the discussion around \eqref{magnetic squark mass}, the $N_f$ squarks are assumed to be tachyonic throughout this region. This is reasonable as one can go to arbitrarily large values of $k$ while keeping $N_f$ fixed. In this regime the gauge sector becomes heavy and decouples from the dynamics. The main contribution to the mass of the squark ($\phi$) comes from the meson multiplet, which is indeed negative. Thus, our main assumption is that this remains true as we move to finite $k$.

Let us then assume that the magnetic squarks condense. In the brane configuration, this corresponds to Higgsing $N_f$ colour D3 branes via reconnection to $N_f$ flavour D3 branes. This is the Higgs mechanism in the string theory language. The world-volume of the $N_f$ Higgsed D3 branes no longer supports a gauge multiplet as they end on D5s from one side and end on the NS5 brane from the other. However, we still have $\kappa$ colour D3 branes which support a $U(\kappa)_{-k}$ gauge theory with massive gauge field and massive gauginos. The CS mass is still proportional to $k$, and we can integrate out the gauge field and gauginos at energies below $g^2 k$. The reconnection preserves the original $U(N_f)$ global symmetry. We will shortly argue, from the field theory side, that there are $N_f$ scalars in the fundamental after the Higgsing. In the brane set-up these can only come from open strings stretched between the colour branes and $N_f$ Higgsed D3 branes.

Let us try to understand the phenomenon described in the last paragraph in terms of the field theory description of the magnetic theory. Indeed, the Higgsing corresponds to giving a colour-flavour locking vev to the magnetic squark without breaking the global $U(N_f)$. The gauge symmetry breaking pattern is given by
\begin{equation}
U(\kappa+N_f) \rightarrow U(\kappa)\; ,
\end{equation}
leaving the gauginos in the $\Ylinethick{1 pt}\Yboxdim7pt\Yvcentermath1\yng(1,1)$ and $\overline{\Ylinethick{1 pt}\Yboxdim7pt\Yvcentermath1\yng(1,1)}$ of the Higgsed gauge group as well as $N_f$ fundamental squarks. The $N_f$ magnetic quarks become massive due to Yukawa terms. In addition, the meson and the mesino all become massive due to interactions like \eqref{phiMterm} and can be integrated out.
\paragraph{} The IR levels get shifted after integrating out the gaugino according to \eqref{gaugino level-shift magnetic} so that, using the dictionary \eqref{dictionary}, the IR of the magnetic theory in this region of the parameter space is described by 
\begin{equation}\label{magnetic IR}
  \text{magnetic IR:}\quad  U\left(K+\frac{N_f}{2}\right)_{-N_c,-N_c+K+\frac{N_f}{2}}\oplus N_f\; \textnormal{scalars}\;.
\end{equation}
Such a bosonic dual is described in the IR by a Lagrangian that contains, in addition to a CS term with appropriate levels and coupling between the scalars and gauge field, also self-interactions for the squarks. These correspond to mass terms of the form $\bar{\phi}^{a}_i \phi_{a}^i$ as well as quartic interaction of the form (single-trace) $(\bar{\phi}^{a}_i \phi_{a}^j)( \bar{\phi}_{j}^b \phi^{k}_b)$ and (double-trace) $(\bar{\phi}_{ i}^a \phi^{i}_a)^2$. These terms can be generated, if not already present, by the RG flow consistently with global symmetries.

As a final step, tuning the mass terms both in the electric IR theory in \eqref{electric IR} and in the magnetic IR theory in \eqref{magnetic IR}, we recover a well-established duality. This is nothing but the duality \eqref{bosonization}. 

 \subsection{Symmetry breaking}

\paragraph{} When $N^\star>N_f>\kappa $, which corresponds to region II and II$'$ in the phase diagram of figure \ref{phase diagram}, we expect rather different dynamics for the system and we anticipate breaking of the flavour symmetry. As we shall see, the physics in these regions is still captured by a tachyonic squark, colour-flavour locking and brane reconnection, but the implications and the resulting physics will be different with respect to region I. Note that the electric theory we discuss is a $U(N_c)$ gauge theory, while the result in ref.\cite{Komargodski-Seiberg} is for $SU(N_c)$.

\subsubsection{Region II$'$}
\paragraph{} Let us begin with region II' in the phase diagram of figure \ref{phase diagram}. In this region $\kappa<0$. Therefore, on the magnetic side, there are less colour D3 branes than flavour D3 branes: $\tilde{N}_c = N_f + \kappa < N_f$. We will assume that the squarks condense also in this case. Nonetheless, squark condensation leads in this case to a fully Higgsed gauge group. Once again this is realised in string theory by reconnecting $N_f+\kappa$ colour and flavour D3 branes (we stress that $\kappa<0$ here). After the Higgsing, we are left with $|\kappa|$ flavour D3 branes stretched between the D5 brane and the $(1,k)$ fivebrane, as well as the $N_f+\kappa$ connected D3 branes. The latter no longer support a gauge multiplet and therefore gauge symmetry is fully broken.

The global symmetry now consists of a $U(N_f+\kappa)$ factor corresponding to the symmetry on the $N_f+\kappa$ reconnected branes as well as a $U(\kappa)$ factor from the remaining flavour D3 branes. Using the dictionary \eqref{dictionary} we have that in this region the global symmetry breaking pattern is
\begin{equation}\label{symmetry breaking pattern}
    SU(N_f)\rightarrow S \left[ U\left(\frac{N_f}{2}+K\right)\times U\left(\frac{N_f}{2}-K\right) \right]\;.
\end{equation}
This symmetry breaking pattern is the one anticipated in \cite{Komargodski-Seiberg}. As a consequence, the IR physics of this phase is described in terms of the Grassmannian
\begin{equation}
\mathcal{M} \left(K + \frac{N_f}{2}, N_f \right) = \frac{SU(N_f)}{S \left[U\left(\frac{N_f}{2}+K\right)\times U\left(\frac{N_f}{2}-K\right) \right]}
\end{equation}
corresponding to the symmetry breaking pattern given in \eqref{symmetry breaking pattern}. Such a Grassmannian will be essentially parametrised by\footnote{In order to be consistent with the UV symmetries one must also include CS terms in the effective description. The required modification is discussed in detail in \cite{Komargodski-Seiberg}.  }
\begin{equation}
    N_f^2 - 1 - \left[(N_f + \kappa)^2 + \kappa^2 - 1\right] = 2|\kappa|(N_f - |\kappa|)=2\left(\frac{N_f}{2} + K\right) \left(\frac{N_f}{2} - K \right) 
\end{equation}
massless Nambu-Goldstone bosons. We identify the Nambu-Goldstone bosons as the massless modes of open strings stretched between the two stacks of flavour branes.

\subsubsection{Region II}\label{region II}
\paragraph{} When $0<\kappa<N_f<N^\star$ (or  $0<K+\frac{N_f}{2}<N_f<N^\star$), after reconnection the theory in the IR is 
\begin{equation}\label{region 2 bosonic dual}
    U\left (K+\frac{N_f}{2} \right )_{-N_c,-N_c+K+\frac{N_f}{2}}\oplus N_f \;\phi\;.
\end{equation}
Naively, we seem to have a puzzle: instead of obtaining a theory of massless Nambu-Goldstone bosons we obtain bosonization. The NG theory we are seeking is nothing but the effective description of \eqref{region 2 bosonic dual} for large negative masses of the squarks $\phi$. According to the field theory analysis of Komargodski and Seiberg \cite{Komargodski-Seiberg} upon condensation of the squarks we land on the symmetry breaking phase.

Indeed, after reconnection, the scalars in the bosonic dual \eqref{region 2 bosonic dual} correspond to scalar modes of the open strings in the brane configuration. Therefore our proposal is that these scalars are tachyonic and are to be stabilised via open string tachyon condensation. We do not know whether a nice geometric picture emerges after this condensation. Regardless, in the field theory limit one eventually lands on the Grassmannian $\mathcal{M}(N_f,\kappa)$. This picture is consistent with the mass deformations of the brane setup, already discussed in \cite{Armoni:2017jkl}.

\section{Comments about QED$_3$}
\paragraph{} The discussion of the phase diagram in the preceding sections holds for a general number of colours $N_c$. However, ``accidents'' happen when $N_c=1,2$ that modify parts of the discussion. In the case of $N_c=2$ the electric gaugino is a singlet of the $SU(2)$ factor of the gauge group, but it carries charge $2$ under the abelian factor. Because of this, some intermediate steps taken to arrive at the general phase diagram in figure \ref{phase diagram} are slightly modified, the end result is however unaffected and the phase diagram of figure \ref{phase diagram} is the correct picture for $N_c\geq2$.

On the other hand, we start to see deviations from the general picture of figure \ref{phase diagram} for $N_c=1$ i.e. QED$_3$. In particular, as we shall see momentarily, when $k=0$ there is no symmetry breaking phase. This in turn suggests that no symmetry breaking can occur for non-zero $k$ since the window for which a Grassmannian phase exists in the IR is maximised for $k=0$ \cite{Komargodski-Seiberg}. 
\subsection{QED$_3$ with vanishing CS-term}
\paragraph{} When the electric gauge group is $U(1)$, there is no electric gaugino. Therefore, the IR of the electric theory is $U(1)_0$ theory coupled to $N_f$ fermions. The magnetic dual has a gauge group $U(N_f+1)$ with vanishing CS level at tree-level. Previously, squark condensation lead to masses being generated for the quarks, meson and the mesino, due to the presence of Yukawa interactions. However, in this case after reconnection we have a $U(1)$ gauge theory with no CS term and $N_f$ massless Dirac fermions. The reason that in this specific case the fermions do not acquire a mass is that there is no gluino when the gauge group is $U(1)$ and no Yukawa term. In the absence of supersymmetry and without fine-tuning the squarks acquire a mass. So we end up with a magnetic theory that admits the same matter content as the electric theory, namely a dual $U(1)$ theory with $N_f$ dual quarks. 

The brane setup is such that the flavour branes coincide and hence flavour symmetry remains unbroken. Thus, our magnetic theory predicts no spontaneous breaking of $U(N_f)$. This is consistent with existing conjectures about the IR behaviour of QED$_3$ \cite{Cordova:2017kue}.  

\section{Conclusions}
\paragraph{}In this manuscript we discussed QCD$_3$ based on a unitary group and its embedding in string theory. The UV field theory on the brane configuration consists of fields that acquire a mass and decouple as the theory flows to the IR. The advantage of having such a UV theory is that it admits a Seiberg duality. The magnetic Seiberg dual leads to new insights about QCD$_3$. In particular the bosonized theory admits a simple realisation as a magnetic dual of the electric fermionic theory. While in the electric side scalar quarks acquire a mass and decouple, in the magnetic side the fermionic quarks acquire a mass due to Yukawa coupling and decouple. 

The Seiberg dual also enables us to gain a better understanding of the symmetry breaking phase. Triggered by condensation of the dual squark the magnetic gauge theory is completely Higgsed and flavour symmetry gets broken.

In addition, we learned about the abelian theory, with or without a Chern-Simons term. The level $k$ (with $k \geq 0$) $U(1)$ theory with $N_f$ flavours admits a magnetic dual that upon Higgsing flows to another $U(1)$ theory with $k'=-k$ and $N_f$ flavours. Flavour symmetry is not broken, as expected from field theory analysis. For $k=0$ the theory looks self-dual. While for $N_f=2$ the self duality is well understood \cite{Komargodski-Seiberg}, for $N_f \ne 2$ the naive self-duality deserves further investigation.

We haven't discussed the regime of $N_f > N^\star$. This regime is hard to analyse both in field theory and in string theory. As in the symplectic case \cite{Armoni:2017jkl} we anticipate that it is described by meson condensation.

\vskip 1cm

{\bf Acknowledgement} We thank Vasilis Niarchos for numerous useful discussions and Zohar Komargodski for a careful reading of the manuscript. We are grateful to Dan Thompson for pointing out a typographical error in the draft.   The work of A.A. has been supported by STFC grant ST/P00055X/1.

  \bibliography{Bibliography13}
\end{document}